\documentclass[aps,prb,reprint,superscriptaddress,floatfix]{revtex4-1}
\usepackage{xcolor}
\usepackage{amsmath,bm,epsfig}
\definecolor{g-blue}{rgb}{0.83,0.95,1}
\usepackage{bm}
\usepackage{amssymb}
\usepackage{amsfonts}
\usepackage{amsmath}
\usepackage{graphicx}

\usepackage[normalem]{ulem}  % to strike out text with \sout{text}
\usepackage[pdftex,colorlinks=true,allcolors=blue,breaklinks=true]{hyperref}  %the build LaTeX => PDF
\makeatletter
\def\set@curr@file#1{%
	\begingroup
	\escapechar\m@ne
	\xdef\@curr@file{\expandafter\string\csname #1\endcsname}%
	\endgroup
}
\def\quote@name#1{"\quote@@name#1\@gobble""}
\def\quote@@name#1"{#1\quote@@name}
\def\unquote@name#1{\quote@@name#1\@gobble"}
\makeatother
\newcommand{\eq}[1]{(\ref{#1})}%%  requires \eq{label}
\newcommand{\Eq}[1]{Eq.\,(\ref{#1})}%%  requires \eq{label}
\newcommand{\Eqs}[1]{Eqs.\,(\ref{#1})}%%  requires \eq{label}
\newcommand{\Fig}[1]{Fig.\,\ref{#1}}%%  requires \Fef{label}
\newcommand{\Sec}[1]{Sec.\,\ref{#1}}%%  requires \Fef{label}
\newcommand{\Refn}[1]{Ref.\,\cite{#1}}%%  requires \Fef{label}
 \def\Fbox#1{\vskip1ex\hbox to 8.5cm{\hfil\fboxsep0.3cm\fbox{%
			\parbox{8.0cm}{#1}}\hfil}\vskip1ex\noindent}  %%  {TEXT} 

\renewcommand{\sb}[1]{_{\text {#1}}}  %% sub-   for lower case
\renewcommand{\sp}[1]{^{\text {#1}}}  %% super- for lower case
\newcommand{\Sp}[1]{^{^{\text {#1}}}} %% Super- for Upper case

\def\He4 {$^4$He~}
\newcommand{\B}[1]{{\bm{#1}}}%% Bold Roman & Greek Lower & Upper Case
\newcommand{\C}[1]{{\mathcal{#1}}}%%  Calligraphic Upper case

\begin{document}
%\linenumbers
\title{Double accumulation and anisotropic transport of magneto-elastic bosons \\ in yttrium iron garnet films}
	
	\author{Pascal~Frey}
	\email{pfrey@rhrk.uni-kl.de}
	\affiliation{Fachbereich Physik and Landesforschungszentrum OPTIMAS, Technische Universit\"at Kaiserslautern, 67663 Kaiserslautern, Germany}
	
	\author{Dmytro~A.~Bozhko}
	\email{dbozhko@uccs.edu}
	%\affiliation{School of Engineering, University of Glasgow, Glasgow G12 8LT, United Kingdom}
	\affiliation{Department of Physics and Energy Science, University of Colorado at Colorado Springs, Colorado Springs CO 80918, USA}
	
	\author{Victor~S.~L'vov}
	\email{victor.lvov@gmail.com}
	\affiliation{Department of Chemical and Biological Physics, Weizmann Institute of Science, Rehovot 76100, Israel}
	
	\author{Burkard~Hillebrands}
	\email{hilleb@physik.uni-kl.de}
	\affiliation{Fachbereich Physik and Landesforschungszentrum OPTIMAS, Technische Universit\"at Kaiserslautern, 67663 Kaiserslautern, Germany}
	
	\author{Alexander~A.~Serga}
	\email{serga@physik.uni-kl.de}
	\affiliation{Fachbereich Physik and Landesforschungszentrum OPTIMAS, Technische Universit\"at Kaiserslautern, 67663 Kaiserslautern, Germany}
	
	%\date{\today}	
	
	\begin{abstract}
 Interaction between quasiparticles of a different nature, such as magnons and phonons in a magnetic medium, leads to the mixing of their properties and the formation of hybrid states in the areas of intersection of individual spectral branches. We recently reported the discovery of a new phenomenon mediated by the magnon-phonon interaction: the spontaneous bottleneck accumulation of magneto-elastic bosons under electromagnetic pumping of pure magnons into a ferrimagnetic yttrium iron garnet film.
 Here, by studying the transport properties of the accumulated magneto-elastic bosons, we reveal that such accumulation occurs in two frequency-distant groups of quasiparticles: quasi-phonons and quasi-magnons. They propagate with different speeds in different directions relative to the magnetization field. 
 The theoretical model we propose qualitatively describes the double accumulation effect, and the analysis of the two-dimensional spectrum of quasiparticles in the hybridization region allows us to determine the wavevectors and frequencies of each of the groups.
	\end{abstract}
	
	\maketitle
	
% 	\tableofcontents 
%================================================================================================= 	
\section*{Introduction}
The physics of quasiparticles constitutes a very prominent research field over the last decade \cite{QP_Zoo}. Especially magnons and phonons---the quanta of spin waves \cite{Dyson1956} and lattice vibrations \cite{Einstein1907}---have been extensively studied within the solid state research domain.
In the early days, these two systems were usually considered to be non-interacting. 
Subsequently, the emphasis shifted to a full description of the solid-state system and the study of related subsystems \cite{Tiersten1964,Kobayashi1973I,Kobayashi1973II,Dransfeld1959,Pomerntz1961,Rezende1969,Rueckriegel2014,Kikkawa2016,Baryakhtar2017}.
The variety of interactions between these two subsystems creates a wide range of applications \cite{Bozhko2020} exploiting the fact that in magnetostrictive materials the mechanical stress affects the magnetization orientation and vice versa \cite{Olabi2008,Chumak2010,Domann2015,Kamra2015,Guerreiro2015,Kryshtal2017,An2020,Zhao2020}.
For instance, the mechanical stress produced by an acoustic wave can drive magnetization dynamics and excite magnons in such materials if the frequency of lattice vibrations matches the eigenexcitations of the spin system \cite{Geilen2020,Weiler2012}.
Furthermore, hybrid waves \cite{Kittel1958,Strauss1965,Camley1978,Camley1979} with mixed magnonic and phononic features \cite{Kamra2014,Bauer2015,Ogawa2015,Hashimoto2017,Weiler2020} also exist. The corresponding quasiparticles are called magneto-elastic bosons \cite{Bozhko2017} or magnon polarons \cite{Kikkawa2016,Sukhanov2019,Yahiro2020}.
Due to their mixed hybrid state between phonons and magnons, they are capable of carrying spin information at velocities close to those of phonons. This is currently attracting much attention to these quasiparticles as promising data carriers for future applications in spintronics \cite{Kamra2014,Ogawa2015,Flebus2017,Ruckriegel2020,Holanda2021}.
In particular, the spontaneous bottleneck accumulation of hybrid magneto-elastic quasiparticles with rather high group velocities was recently discovered in an overpopulated magnon gas in yttrium iron garnet (YIG) films \cite{Bozhko2017}.  

\begin{figure} [t]
	\includegraphics[width=8.6cm]{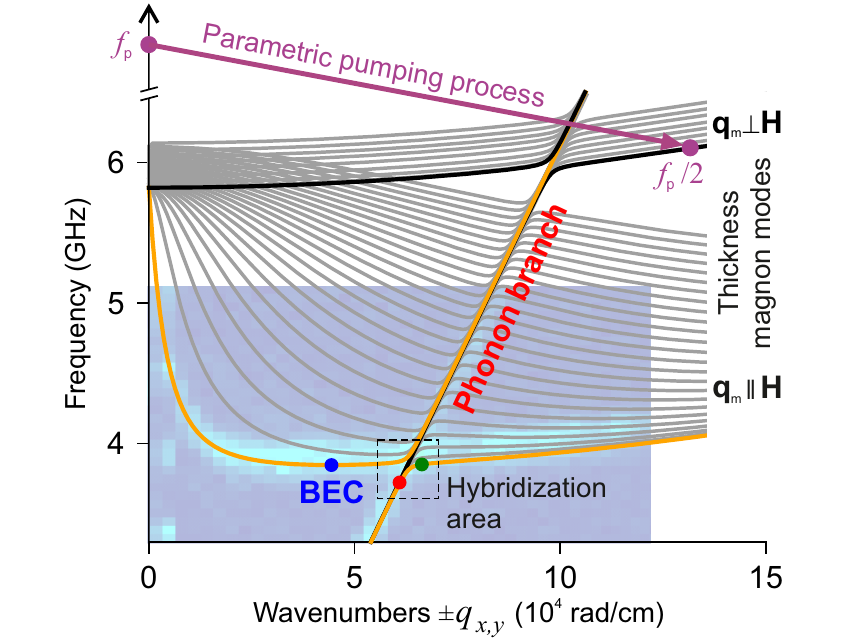}
	\caption{
		\label{dispersion} 
	 A magnon-phonon hybrid spectrum of an in-plane magnetized 5.6\,$\mu$m-thick YIG film calculated for wavevectors $\pm q_\mathrm{x}$ and $\pm q_\mathrm{y}$ oriented along and perpendicular to the bias magnetic field, respectively. The external magnetic field is $\mu_0 H = $135\,mT. The violet arrow indicates the parallel parametric pumping process in the $\textbf{\textit{q}}_\mathrm{m} \perp \textbf{\textit{H}}$ magnon branch using microwave photons of frequency $f_\mathrm{p}$. The population of the pumped magnons at frequency $f_\mathrm{p}/2$ (purple dot)  thermalizes over the spin-wave spectrum via four-magnon scattering processes. The thermalization process leads to the accumulation of magneto-elastic quasiparticles within the magnon-phonon hybridization area (dashed square) and magnon Bose-Einstein condensation at the bottom of the spin-wave spectra (blue dot). The green and red dots mark the spectral positions of the measured hybrid quasiparticles. The color map shows the thermal magnon-phonon density distribution measured by means of frequency- and wavevector-resolved Brillouin light scattering spectroscopy.
	 }
\end{figure} 

In this paper, we report measurements of the two-dimensional transport properties of the accumulated quasiparticles using space- and time-resolved Brillouin light scattering (BLS) spectroscopy. We show that the accumulation process leads to the appearance of two---slow and fast---groups of quasiparticles, represented by the green and red dots in the black dashed square in Fig.\,\ref{dispersion}, propagating in distinctly different directions. The slow group, indicated by the red dot, has a frequency slightly higher than that of the magnon Bose-Einstein condensate (BEC) \cite{Demokritov2006,Serga2014} formed at the bottom of the spin-wave spectrum. The frequency of the fast group of hybrid bosons, shown by the red dot, is lower then the BEC frequency. 

We also present an analytical model that accounts for the details of the relaxation dynamics in the presence of a nonzero BEC frequency minimum including variation of the scattering coefficient in the region of the avoided crossing between phonon and magnon branches. The model qualitatively explains the observed phenomenon of double accumulation and the spectral positions of the two quasiparticle groups. 

Our analysis of the two-dimensional hybrid frequency spectra quantitatively relates the values and directions of quasiparticle velocities to the formation of caustic beams \cite{Veerakumar2006,Serga2010} in the spatial distribution of the quasiparticle density. The presented findings will potentially allow to control the characteristics of the accumulated quasiparticles and manipulate their behavior for information transfer and processing \cite{Holanda2018,Heussner2018,Hioki2020}.

The paper is organized as follows. The experimental setup is described in \Sec{ss:setup}. In \Sec{ss:results}, the experimental results on propagation velocities of the slow and fast quasiparticle packets are presented.  In \Sec{s:background}, we refer to the basics of magnon-phonon hybridization theory by deriving a hybridized magnon-phonon Hamiltonian in \Sec{ss:hybrid} and presenting a statistical description of magneto-elastic modes in \Sec{ss:stat}. The model of double accumulation of hybrid bosons is presented in \Sec{s:model}, where we introduce a one-dimensional differential model in \Sec{ss:dif}, estimate the scattering rates within and between the quasiparticle spectral branches in \Sec{ss:scattering}, and solve the rate equations above and below the BEC frequency in \Sec{ss:sol-1} and \Sec{ss:sol-2}, respectively. Analysis of the experimental data based on the theory of caustic magnon transport in \Sec{s:2D} allows us to determine frequencies and wavevectors of the two groups of hybrid quasiparticles. In \Sec{s:discussion}, we discuss and summarize the obtained results. 

%=================================================================================================
\vspace{-3mm}
\section{Experiment}
\vspace{-3mm}
\subsection{\label{ss:setup} Setup}
\vspace{-3mm}

\begin{figure}[b]
	\includegraphics[width=8.6 cm]{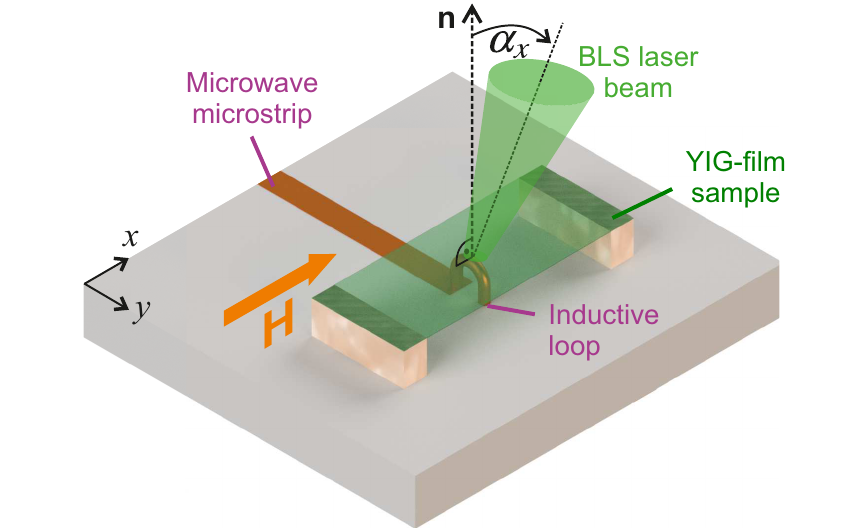}
	\caption{
		\label{section} 
		Sketch of the experimental setup. The microwave part consist of a microwave stripline, which blends into a half-loop with its plane oriented perpendicular to the sample surface and the external magnetic field $\textbf{\textit{H}}$. The sample is mounted on top of two distant holders in such a way that the center of the sample is located above the excitation loop. The sample itself consists of a YIG film of 5.6\,$\mathrm{\mu m}$ thickness grown on a GGG substrate (500\,$\mathrm{\mu m}$) and covered by a thin dielectric mirror coating ($< 1 \,\mathrm{\mu m}$). The magnons in the YIG film are detected via inelastic laser light scattering ($\lambda=532$\,nm) measured by a tandem Fabry-P\'{e}rot interferometer. The value of an incident angle $\alpha_x$ of the probing laser beam in the $(x,z)$-plane defines the measured wavevector $q_x$ oriented along the magnetic field.
		}
\end{figure}

% description of experimental setup
We studied the process of spontaneous accumulation of hybrid quasiparticles and determined their properties by means of BLS spectroscopy \cite{Sandercock}. The sample is illuminated by focused laser light with a wavelength of 532\,nm. Due to inelastic scattering processes, a photon can absorb or excite a magnon and therefore gain (anti-Stokes process) or lose energy (Stokes process). Thus, the intensities of the Stokes and anti-Stokes spectral peaks are proportional to the density of quasiparticles. The backward scattered light is collected and directed to a tandem Fabry-P\'{e}rot interferometer, where its frequency spectrum is analyzed with a resolution of about 100\,MHz. By carefully selecting the incident angle $\alpha_x$ of the laser light along the external magnetic field we achieve a wavevector selection (see Fig.\,\ref{section}) \cite{Sandweg2010,Bozhko2020_unconventional}. The linear response function of the BLS setup in the wavevector plane $(q_x, q_y)$ follows a Gaussian distribution and consequently limits the wavevector resolution with a standard deviation of $\sigma$ = 1500\,rad/cm. To freely move the sample along all three spatial dimensions it is mounted on a motorized linear stage system. Additionally, a time-resolution system with a resolution of 400\,ps is combined with the interferometer and a pulsed microwave source \cite{Buettner2000}. Therefore, the setup is able to perform \mbox{frequency-,} \mbox{wavevector-,} \mbox{time-,} and space-resolved scans \cite{THATec}. 

 \begin{figure*}
	\includegraphics[width=1\textwidth]{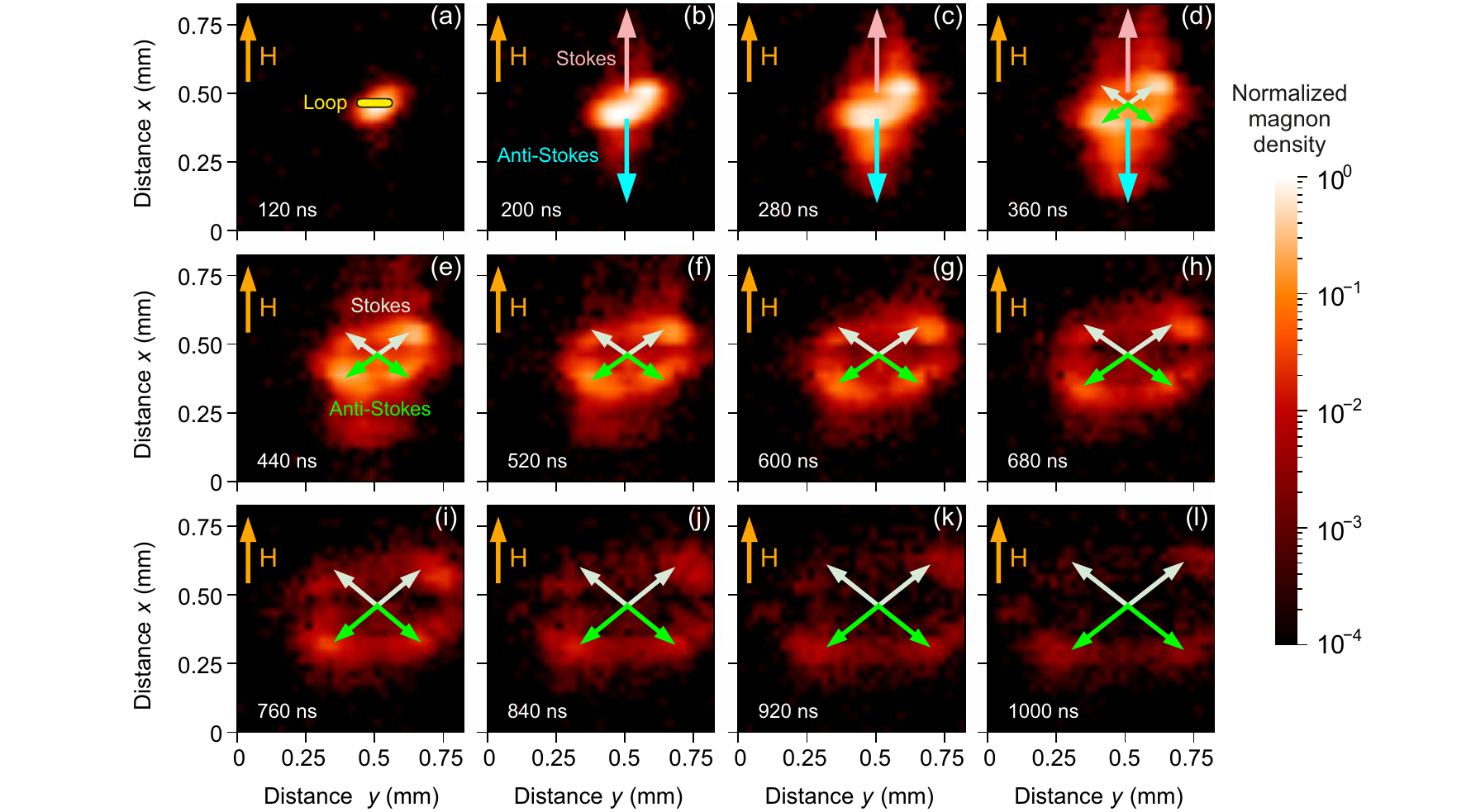}
	\caption{
		\label{timeshoots}  
		Spatial distribution of the BLS-signal intensity corresponding to the distribution of the quasiparticle density for different times from the moment in time the pump pulse is turned on. The propagation of two groups of quasiparticles is visible. The magnon-phonon hybrid packets propagate away from the excitation loop in six directions. Each of the shown intensity distributions are a summation of the BLS experiment's Stokes ($+q_x$) and anti-Stokes ($-q_x$) signals.
		}
\end{figure*}  

%description of the sample structure 
The measurements are carried out at room temperature in a single-crystal Yttrium Iron Garnet (YIG: Y$_3$Fe$_5$O$_{12}$) sample. This ferrimagnetic material was chosen for the experiment because it uniquely combines extremely low damping of magnetic and elastic excitations \cite{Cherepanov1993}. The sample is a piece of 5.6\,$\mu$m-thick YIG film grown by liquid phase epitaxy on top of a 500\,${\mu}$m-thick Gadolinium Gallium Garnet (GGG: Gd$_3$Ga$_5$O$_{12}$) substrate in (111) crystallographic plane (see Fig.\,\ref{section}). Micron-thick YIG films as well as GGG substrates are transparent to green light and thus make BLS probing of magnon-phonon hybrid waves possible. The probing light beam is focused onto the YIG film through the GGG substrate. To ensure homogeneous reflection of the probing light from the YIG-film surface and thus uniform spatial sensitivity of the BLS setup, YIG is covered with a thin dielectric mirror coating created with a SiO$_2$/TiO$_2$ bi-layer structure ($<$1\,$\rm{\mu m}$). The mirror reflects over 90\,$\%$ light at 532\,nm.

% description of magnon injection 
Magnons are excited in the in-plane magnetized YIG sample by microwave electromagnetic pumping via the parallel parametric pumping process. In such a process \cite{Lvov1994}, one microwave photon of the pumping frequency $f_\text{p}$ splits into two magnons with opposite wavevectors $\pm \textbf{\textit{q}}_\mathrm{m}$ and each the frequency $f_\text{p}/2$.  
Due to the choice of the values of the bias magnetic field $\mu_0 H = 135\,$mT and the pumping frequency $f_\text{p}=14\,$GHz, the parametrically excited magnons are injected (see violet arrow in Fig.\,\ref{dispersion}) into the transverse branch ($\textbf{\textit{q}}_\mathrm{m} \perp \textbf{\textit{H}}$) of the spin-wave frequency spectrum \cite{Serga2012} at a distance of about 2\,GHz above the bottom of the spectrum.

In this experiment, to achieve two-dimensional spatial localization of the pumping microwave field required for two-dimensional transport measurements, the microstrip pump resonator used in our previous experiments \cite{Bozhko2017,Bozhko2018} is replaced by a half-loop antenna with a diameter of 100\,$\mu$m (see Fig.\,\ref{section}). The half-loop is made of a gold wire with a diameter of 25\,$\mu$m and is shorted to the ground of the microstrip line that feeds the pumping power to it. 
Since the quality factor of this antenna is unity, the pumping power required to reach a pronounced accumulation of hybrid quasiparticles and the Bose-Einstein condensation of magnons is about 400\,W.
This power is applied during a 200\,ns-long excitation pulse. To ensure that no additional heating effects influence the measurements, the excitation pulse is followed by a 250\,$\mu$s idle time, where the magnons can decay and the created heat can dissipate into the surroundings.

 \begin{figure}
	\includegraphics[width=8.6 cm]{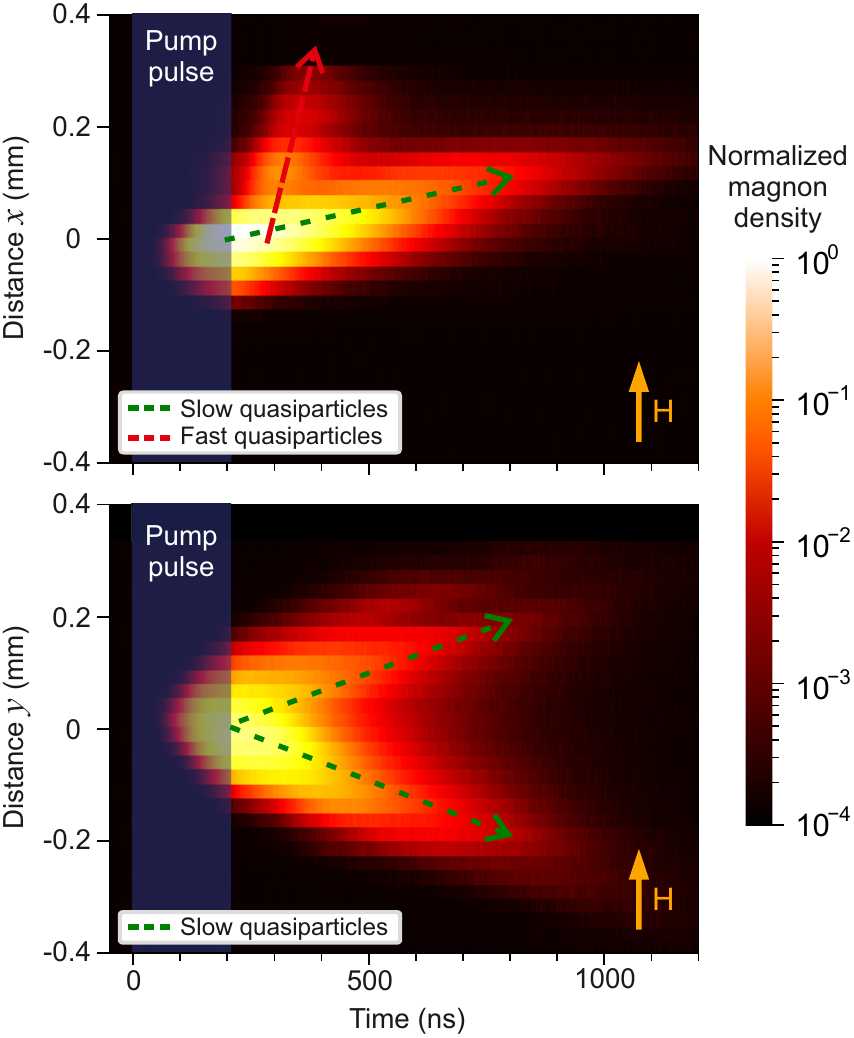}
	\caption{
		\label{space-time_diagrams} Space-time diagrams along the $x$- and $y$-axes in the film plane for packets of fast and slow quasiparticles, which correspond to Stokes components of the BLS signals and thus have positive values of wavevectors $q_x$. The green and red dashed lines show a linear temporal fit of the positions of these packets. 
		 Two packets of quasiparticles with wavevectors $\pm q_y$ propagate with the same speed in opposite directions along the $y$ axis, and two packets -- along the $x$ axis with different speeds.}
\end{figure} 

%-------------------------------------------------------------------------------------------------
\vspace{-3mm}
\subsection{\label{ss:results}Experimental results}
\vspace{-3mm}
The measurement cycle begins at $t=0$, when the pumping power is switched on and when a microwave pulse begins to inject magnons into the system. From thereon the magnon gas is formed and it thermalizes via four-magnon scattering processes \cite{NSW,Serga2014}. 
The thermalization process causes an increase in the local chemical potential of the injected magnon gas in the vicinity of the frequency minimum. As soon as it reaches the minimum of the magnon frequency, a magnon BEC forms at the spot marked by the blue dot in Fig.\,\ref{dispersion}. 

Another cluster of quasiparticles is formed at the intersection of the phonon and magnon branches \cite{Bozhko2017}. Quasiparticles belonging to this cluster are in hybrid states between phonons and magnons, combining the properties of both. Due to the rather small size of the hybridization region in the frequency and wavevector space, the BLS setup is not able to differentiate the population of quasiparticles within the magneto-elastic interaction region. However, time-resolved measurements of the spatial distribution of these quasiparticles \mbox{[see Figs.\,\ref{timeshoots}(a-l)]} reveal the presence of two groups of quasiparticles with significantly different propagation properties. 

Figure\,\ref{timeshoots}(a) depicts the beginning of quasiparticle accumulation in real space in the vicinity of the pump loop at time $t=120\,$ns after the start of pumping. Here, the loop position is shown by the gold segment. During the pumping pulse ($t\le 200\,$ns) the number of particles in the hybridization region rises [see Figs.\,\ref{timeshoots}(a,\,b)], and two beams of quasi-particles start to travel   upward  and downward along the direction of the external magnetic field $\B H$ as shown in Figs.\,\ref{timeshoots}(b,\,c). The white arrows mark the Stokes data, indicating the quasiparticles propagating with a positive wavevector of $q =q\sb{fast}$, while the blue arrows for anti-Stokes data correspond to the quasi particles propagating with a negative wavevector $q=-q\sb{fast}$.

Beginning at $t=360$\,ns [see Fig.\,\ref{timeshoots}(d)], four additional magnon packets propagate obliquely to the external magnetic field as it is marked by two white and two green arrows in Figs.\,\ref{timeshoots}(e-l)]. The speed of their propagation is approximately ten times lower than the speed of propagation of quasiparticle packets along the field. Hereinafter we will refer to these two groups of quasiparticles as fast and slow ones.

To analyze the propagation process of the fast and slow packets, a center-of-mass algorithm tracks the path of each of those packets individually. The space-time diagrams for the fast and slow packets in $x$- and $y$-directions are shown in Fig.\,\ref{space-time_diagrams}(a) and Fig.\,\ref{space-time_diagrams}(b).
The packets positions $\B R(t)=\{X(t), Y(t)\}$, represented by the red and green arrows, are found by the two-dimensional linear in time fitting process $\B R(t)= \B v t + \B R_0$ with the Levenberg-Marquardt algorithm \cite{More1977}:
\begin{subequations}\label{V} 
\begin{eqnarray}\label{slow} 
v\sb{slow}&\approx& (368\pm 2)  \mbox{m/s}, \qquad \alpha\sb{slow}\approx 59^\circ\pm 0.3^\circ\,,\\
 \label{fast}
v\sb{fast}&\approx& (3070\pm 207)  \mbox{m/s}, \ \  \alpha\sb{fast}\approx 0^\circ\pm 0^\circ\ .
\end{eqnarray}
\end{subequations}
%
%The confidence intervals are determined for the confidence probability of $95\%$ using experimental Gaussian errors calculated from the covariance matrix. 

We see that the velocity of the slow quasiparticle group is higher than that of pure magnons ($\approx 90$\,m/s) with wavevectors corresponding to the hybridization region, while the velocity of the fast group is lower than that of pure phonons ($\approx 3850$\,m/s in bulk YIG). This comparison of velocity values allows us to approximately localize the spectral positions of the fast and slow quasiparticles in the hybridization region--indicated by the red and green dots inside the black dashed box in Fig.\,\ref{dispersion}. It can be assumed that the slow quasiparticles are mainly of magnonic nature with a small contribution of phononic admixture. In turn, the fast quasiparticles predominantly consist of phonons with a small magnonic contribution. Particularly remarkable is the absence of the population of quasiparticle states between these two---quasi-magnon and quasi-phonon---extreme cases. Because of close spectral positions of the quasi-magnon and quasi-phonon states, such selective population is difficult to explain by some features of scattering of parametrically pumped magnons far from the bottom of the spin-wave spectrum. Thus, to understand the observed phenomenon of the double quasiparticle accumulation, a theoretical analysis of the situation near the magnon-phonon hybridization region is required. It is presented in the following sections.

%================================================================================================
 \section{\label{s:background} Magnon-phonon hybridization background}
 \vspace{-3mm}
 
 The goal of this section is to recall the Hamiltonian description of interacting magnons and photons in the vicinity of cross-section of their frequency spectra\,\cite{ZLF,NSW}, required for further discussion of our experimental finding. Some details on statistical descriptions of resulting magneto-elastic modes can be found in the Appendix of \Refn{arc-1}.

%-------------------------------------------------------------------------------------------------
\vspace{-3mm}
\subsection{\label{ss:hybrid}Hybridized magnon-phonon Hamiltonian}  
\vspace{-3mm}

At high temperatures we can restrict ourselves to the classical limit and describe the system of interacting magnons and phonons in the framework of a classical Hamiltonian formalism. This approach is applicable to a wide class of weakly interacting wave systems, allowing a physically transparent and very compact description of their common properties, see, for instance, Chapter\,1 in \Refn{NSW} and in more detail in \Refn{ZLF}. 
 
Introducing complex canonical amplitudes of magnons and phonons in the wavevector $\textbf{\textit{q}}$ representation, $a(\textbf{\textit{q}},t) \equiv a_\textbf{\textit{q}}$ and $b(\textbf{\textit{q}},t) \equiv b_\textbf{\textit{q}}$ we can write their Hamiltonian equation of motion as follows:
 \begin{subequations}\label{Ham}
 	\begin{equation}\label{HamA}
 	i\, \frac{\partial a_\textbf{\textit{q}}}{\partial t}=\frac{\partial \mathcal{H}}{\partial a^*_{\textbf{\textit{q}}}}\,, \quad i\, \frac{\partial b_\textbf{\textit{q}}}{\partial t}=\frac{\partial \mathcal{H}}{\partial b^*_\textbf{\textit{q}}}\, \ .
 	\end{equation}
 	Here, the Hamiltonian function $\mathcal{H}$ (hereafter referred to as``the Hamiltonian'' for brevity) is a functional of the canonical variables $a_\textbf{\textit{q}}$, $b_\textbf{\textit{q}}$ for all $\B q$ and their complex conjugated counterparts  $a^*_\textbf{\textit{q}}$, $b^*_\textbf{\textit{q}} $. We chose $\mathcal{H}$ as
 	\begin{eqnarray}\label{HamB}
 	\mathcal{H} &=& \mathcal{H}_2+ \mathcal{H}_4\,, \\ 
 	\mathcal{H}_2&=&\sum_\textbf{\textit{q}}\Big[\omega^\textrm{m}_\textbf{\textit{q}}a_\textbf{\textit{q}}a^*_\textbf{\textit{q}}+ \omega^\textrm{p}_\textbf{\textit{q}}b_\textbf{\textit{q}}b^* _\textbf{\textit{q}} +\frac{\Delta}2 \Big (a_\textbf{\textit{q}}b^*_\textbf{\textit{q}}+a^*_\textbf{\textit{q}}b_\textbf{\textit{q}} \Big ) \Big]\,, ~~\label{HamC}\\
 	\mathcal{H}_4&=&\frac{1}{4} \sum_{\textbf{\textit{q}}_1+ \textbf{\textit{q}}_2=\textbf{\textit{q}}_3+\textbf{\textit{q}}_4}T_{\textbf{1} \textbf{2},\textbf{3} \textbf{4}}\,a_\textbf{1}^* a^*_\textbf{2} 
 	a_\textbf{3}a_\textbf{4}\,, \quad a_{\B j}\equiv  a_{\B q_j} .~~~ \label{HamD}
 	\end{eqnarray}
 \end{subequations}
 The first two terms in $\mathcal{H}_2$  describe the free propagation of magnons and phonons with the dispersion laws $\omega^\textrm{m}_\textbf{\textit{q}}$ and $\omega^\textrm{p}_\textbf{\textit{q}}$,respectively. The last two terms in $\mathcal{H}_2$ are responsible for their linear coupling due to the magneto-elastic effect with a coupling amplitude $\Delta$. In the acoustic system nonlinearity can be safely neglected in comparison with the strongly nonlinear spin-wave (magnon) subsystem.
 
 Without interaction, the dispersion laws $\omega^\textrm{m}_\textbf{\textit{q}}$ and $\omega^\textrm{p}_\textbf{\textit{q}}$ will cross at some hybridization wavevector $\textbf{\textit{q}}=\textbf{\textit{q}}_\times: \omega_{\textbf{\textit{q}}_\times}^{\textrm m}=\omega_{\textbf{\textit{q}}_\times}^{\textrm p}=\omega_\times$, as seen in Fig.\,\ref{dispersion}. Here, we consider a weak coupling regime with a narrow hybridization region $\delta q$ around $\textbf{\textit{q}}_\times$, such that
 \begin{equation}
 \Delta\equiv \delta q\frac{\partial (\omega_{\textbf{\textit{q}}_ \times}^{\textrm p}-\omega_{\textbf{\textit{q}} _\times}^{\textrm m})}{\partial q}\ll \omega_\times\ .
 \end{equation}
  
 Being interested in the system evolution in the hybridization region $\delta q$, we can restrict ourselves to the four-magnon interaction Hamiltonian $\mathcal{H}_4$, Eq.\,(\ref{HamD}) \cite{NSW} with the sum restricted by the hyper-surface $\textbf{\textit{q}}_1+ \textbf{\textit{q}}_2=\textbf{\textit{q}}_3+\textbf{\textit{q}}_4$.  The interaction amplitude $T_{\textbf{1} \textbf{2},\textbf{3} \textbf{4}}$ depends on $\textbf{\textit{q}}_1$, $\textbf{\textit{q}}_2$, $\textbf{\textit{q}}_3$, $\textbf{\textit{q}}_4$.
 
 The quadratic Hamiltonian $\mathcal{H}_2$ can be diagonalized by the linear canonical Bogoliubov $(u,v)$-transformation of the form:
 \begin{subequations} \begin{equation}\label{u-v}
 \begin{array}{ccc}
 a_\textbf{\textit{q}}&=& c^{\mbox{\tiny L}}_\textbf{\textit{q}} \cos \varphi_\textbf{\textit{q}} + c^{\mbox{\tiny U}}_\textbf{\textit{q}}\sin \varphi_\textbf{\textit{q}} \,, \\ 
 b_\textbf{\textit{q}}&=& - c^{\mbox{\tiny L}}_\textbf{\textit{q}} \sin \varphi_\textbf{\textit{q}} + c^{\mbox{\tiny U}}_\textbf{\textit{q}} \cos \varphi_\textbf{\textit{q}}\,, 
 \end{array}
 \end{equation}
 in which the coordinate system rotation angle $\varphi_\textbf{\textit{q}}$ is chosen as follows:
 \begin{eqnarray} \label{3b}
 ~\hskip - .7 cm\cos^2 \varphi_{\B q}= \frac 12 \Big[ 1+ \frac {O_{\B q}}{ \big(1+O_{\B q}^2\big)^{1/2}}  \Big] \,, \  O_q=\frac{\omega_{\B q}\sp p-\omega _{\B q}\sp m}{\Delta}\ .
  \end{eqnarray} \end{subequations}
 Here, $O_q$ is the dimensionless frequency distance from the crossover.  This transformation leads to a new diagonal Hamiltonian $ \widetilde{\mathcal{H}}_2$ in terms of new normal canonical amplitudes of the upper (U) and lower (L) magneto-elastic modes (MEMs) $c\sp{\tiny U}_{\B q}$ end $c\sp{\tiny L}_{\B q}$  with frequencies $\Omega^{^{\rm U}}_\textbf{\textit{q}}$ and $\Omega^{^{\rm L}}_\textbf{\textit{q}}$:
 \begin{subequations}\label{newH2}
 	\begin{eqnarray}\label{newH2A}
 	\widetilde{\mathcal{H}}_2&=& \sum_\textbf{\textit{q}}\Big [\Omega^{^{\rm U}}_\textbf{\textit{q}}c^{\mbox{\tiny U}}_\textbf{\textit{q}}c^{\mbox{\tiny U}*}_\textbf{\textit{q}}+ \Omega^{^{\rm L}}_\textbf{\textit{q}}c_\textbf{\textit{q}}^{\mbox{\tiny L}}c^{\mbox{\tiny L}*}_\textbf{\textit{q}}\Big ]\,,\\ \label{Omegas}
	\Omega^{^{\rm U}}_\textbf{\textit{q}}&=&\frac{1}{2}\Big \{\omega^\textrm{m}_\textbf{\textit{q}} + \omega^\textrm{p}_\textbf{\textit{q}} + \sqrt{\big [\omega^\textrm{m}_\textbf{\textit{q}} - \omega^\textrm{p}_\textbf{\textit{q}} \big ]^2 + \Delta^2} \, \Big \}\ , \\ 
 	\Omega^{^{\rm L}}_\textbf{\textit{q}}&=&\frac{1}{2}\Big \{\omega^\textrm{m}_\textbf{\textit{q}} + \omega^\textrm{p}_\textbf{\textit{q}} - \sqrt{\big [\omega^\textrm{m}_\textbf{\textit{q}} - \omega^\textrm{p}_\textbf{\textit{q}} \big ]^2 + \Delta^2} \, \Big \}\ . 	
  	\end{eqnarray}
As usual, the group velocities of these modes are given by \looseness=-1
 \begin{equation}\label{group} 
    \B v_{\B q}^{^{\rm U, L}} =  \frac {d \, \Omega^{^{\rm U, L}} _{\B q}  } {d  {\B q}}\ .
 \end{equation} 	
 \end{subequations}
 The interaction Hamiltonian\,(\ref{HamD}) in these new variables includes terms ${\mathcal{H}}^{^{\rm UU}}_4, {\mathcal{H}}^{^{\rm LL}}_4$, responsible for the interactions within the U- and L-MEMs, respectively, and a term ${\mathcal{H}}^{^{\rm LU}}_4$, that describs the interaction between them:
   \begin{subequations}\label{newHam}
 	\begin{align}\begin{split}  
 	{\mathcal{H}}^{^{\rm UU}}_4
 	&= \frac{1}{4}\sum_{\textbf{\textit{q}}_1+ \textbf{\textit{q}}_2=\textbf{\textit{q}}_3+\textbf{\textit{q}}_4} \!\!\!\!
 	T^{^{\rm UU}}_{\textbf{1} \textbf{2},\textbf{3} \textbf{4}}
 	\,c_{\textbf{1}}^{{\mbox{\tiny U}}*} c^{\mbox{\tiny U}*}_{\textbf{2}}c^{\mbox{\tiny U}}_{\textbf{3}}c^{\mbox{\tiny U}}_{\textbf{4}}\,,
 	 	\\    T^{^{\rm UU}}_{\textbf{1} \textbf{2},\textbf{3} \textbf{4}} &=  T^{^{\rm UU}}_{\textbf{1} \textbf{2},\textbf{3} \textbf{4}} \sin \varphi_{\textbf{1}}\sin \varphi_{\textbf{2}}\sin \varphi_{\textbf{3}}\sin \varphi_{\textbf{4}}\,, \label{newHamA}  
 \end{split}\\
 \begin{split}
 	\mathcal{H}_4^{^{\rm LL}}&= \frac{1}{4} \sum_{\textbf{\textit{q}}_1+ \textbf{\textit{q}}_2=\textbf{\textit{q}}_3+\textbf{\textit{q}}_4}  \!\!\!\!
 	T^{^{\rm LL}}_{\textbf{1}\textbf{2},\textbf{3} \textbf{4}}
 	\,c_{\textbf{1}}^{{\mbox{\tiny L}}*} c^{\mbox{\tiny L}*}_{\textbf{2}}
 	c^{\mbox{\tiny L}}_{\textbf{3}}c^{\mbox{\tiny L}}_{\textbf{4}}\,,\\
    T^{^{\rm LL}}_{\textbf{1} \textbf{2},\textbf{3} \textbf{4}} &=  T^{^{\rm LL}}_{\textbf{1} \textbf{2},\textbf{3} \textbf{4}} \cos \varphi_{\textbf{1}}\cos \varphi_{\textbf{2}}\cos \varphi_{\textbf{3}}\cos \varphi_{\textbf{4}}\,, \label{newHamB} 	\end{split}\\   
    \begin{split}
 	\mathcal{H}_4^{^{\rm LU}}&= \frac 14\sum_{\textbf{\textit{q}}_1+ \textbf{\textit{q}}_2=\textbf{\textit{q}}_3+\textbf{\textit{q}}_4}  \!\!\!\!
 	T^{^{\rm LU}}_{\textbf{1} \textbf{2},\textbf{3} \textbf{4}}\Big [
 	\,c_{\textbf{1}}^{\mbox{\tiny U}*} c^{\mbox{\tiny U}*}_{\textbf{2}}c^{\mbox{\tiny L}}_{\textbf{3}}c^{\mbox{\tiny L}}_{\textbf{4}}+ \mbox{c.c.}\Big ]\,, 
 	\\  T^{^{\rm LU}}_{\textbf{1} \textbf{2},\textbf{3} \textbf{4}} &=  T^{^{\rm LU}}_{\textbf{1} \textbf{2},\textbf{3} \textbf{4}} \sin \varphi_{\textbf{1}}\sin \varphi_{\textbf{2}}\cos \varphi_{\textbf{3}}\cos \varphi_{\textbf{4}}\ . \label{newHamC}
 \end{split}	\end{align}
 \end{subequations}  
 Here, ``c.c.'' stands for complex conjugation. The energy and particle fluxes within the U- and L-MEM branches, are proportional to $|T^{^{\rm UU}}_{\textbf{1} \textbf{2},\textbf{3} \textbf{4}}|^2$ and $|T^{^{\rm LL}}_{\textbf{1} \textbf{2},\textbf{3} \textbf{4}}|^2$, the energy and particle exchange between modes are proportional to  $|T^{^{\rm LU}}_{\textbf{1} \textbf{2},\textbf{3} \textbf{4}}|^2$.
 
To illustrate how these objects depend on the frequency near the hybridization crossover, where $\omega^\textrm{m}_\textbf{\textit{q}}=\omega^\textrm{p}_\textbf{\textit{q}}$, we plot them in Fig.\,\ref{model}(b) with the shorthand notations:
\begin{equation}\label{TT}
 T^{^{\rm UU}}_q \equiv T^{^{\rm UU}}_{\textbf{\textit{q}},\textbf{\textit{q}};\textbf{\textit{q}},\textbf{\textit{q}} }\,, \ T^{^{\rm LL}}_q \equiv T^{^{\rm LL}}_{\textbf{\textit{q}},\textbf{\textit{q}};\textbf{\textit{q}},\textbf{\textit{q}} }\,, \ T^{^{\rm LU}}_q \equiv T^{^{\rm LU}}_{\textbf{\textit{q}},\textbf{\textit{q}};\textbf{\textit{q}},\textbf{\textit{q}} }\,,
\end{equation}
as functions of the dimensionless distance to the crossover $O_q$ defined by Eq.\,\eqref{3b}.
 
%-------------------------------------------------------------------------------------------------
\vspace{-3mm}
\subsection{ \label{ss:stat} Statistical description of magneto-elastic modes}
\vspace{-3mm}

 A statistical description of weakly interacting waves can be obtained \cite{ZLF} in terms of a kinetic equation, shown below for the continuous limit, when the system size $L$ is much larger than the wavelength $2\pi/k$:

 \begin{equation}\label{KE}
 \frac{\partial n^{\mbox{\tiny L }}(\textbf{\textit{q}}, t)}{\partial t} = \mbox{St}^{^{\rm L }}(\textbf{\textit{q}}, t)\,, \quad \frac{\partial n^{\mbox{\tiny U }}(\textbf{\textit{q}}, t)}{\partial t} = \mbox{St}^{^{\rm U }}( \textbf{\textit{q}}, t)\ .
 \end{equation}
 Here  $n^{\mbox{\tiny L}}_\textbf{\textit{q}} =n^{\mbox{\tiny L}}(\textbf{\textit{q}}, t)$ and $n^{\mbox{\tiny U }}_\textbf{\textit{q}} =n^{\mbox{\tiny U}}(\textbf{\textit{q}}, t)$ are the same-time pair correlations of the L-  and U-MEMs, defined by
 $$
 \langle c^{\mbox{\tiny L}}_\textbf{\textit{q}}  c^{\mbox{\tiny L }}_{\textbf{\textit{q}}'}\rangle = \dfrac{4\pi^2}{L^2} \, \delta(\textbf{\textit{q}}-\textbf{\textit{q}}')  n^{\mbox{\tiny L }}_\textbf{\textit{q}}\,, \  
 \langle c^{\mbox{\tiny L }}_\textbf{\textit{q}}  c^{\mbox{\tiny L }}_{\textbf{\textit{q}}'}\rangle =\dfrac{4\pi^2}{L^2} \, \delta(\textbf{\textit{q}}-\textbf{\textit{q}}') n^{\mbox{\tiny L }}_\textbf{\textit{q}}\,,
 $$ 
 where $\langle \dots \rangle$ stands for  the ensemble  averaging. In the classical limit, when the occupation numbers of Bose particles $n_q (\textbf{\textit{q}}, t) \gg 1$, $n(\textbf{\textit{q}}, t)=\hbar n_q (\textbf{\textit{q}}, t)$.
 In what follows we discuss only the relevant L-MEM population $n^{\mbox{\tiny  L }}_{\textbf{\textit{q}}}$.
 
 The collision integral $\mathrm{St}^{^{\rm L}}(\textbf{\textit{q}}, t)$ may be found in various ways~\cite{ZLF}, including the Golden Rule, widely used in quantum mechanics. Accounting for the 2L$\Rightarrow$2L and 2L$\Rightarrow$2U scattering events, we have
\begin{subequations}\label{St}
 \begin{align}
 	\label{StA}
 	\mbox{St}^{^{\rm L}}(\textbf{\textit{q}} ,t)= &  \mbox{St}^{^{\rm LL}}(\textbf{\textit{q}} ,t) + \mbox{St}^{^{\rm LU}}(\textbf{\textit{q}} ,t)\,, \\ 
 	\label{StB}
 \begin{split}
 	\mbox{St}^{^{\rm LL}}(\textbf{\textit{q}} ,t)= & \frac{\pi}{4} \int  d\textbf{\textit{q}}_1 d\textbf{\textit{q}}_2 d\textbf{\textit{q}}_3 \, \delta(\textbf{\textit{q}}+\textbf{\textit{q}}_1-\textbf{\textit{q}}_2-\textbf{\textit{q}}_3)\\  
 &\times 	 \delta (\Omega_\textbf{\textit{q}}^{^{\rm L}} +\Omega_1^{^{\rm L}}-\Omega_2^{^{\rm L}} -\Omega_3^{^{\rm L}})  \,   
 	|T^{^{\rm LL}}_{\textbf{\textit{q}} 1\, \textbf{2} \textbf{3}}|^2  \, \\  
 &	\times  [n_\textbf{2}^{\mbox{\tiny L}}  n_\textbf{3}^{\mbox{\tiny L}} (n_\textbf{\textit{q}} ^{\mbox{\tiny L}}  + n_\textbf{1}^{\mbox{\tiny L}} )- n_\textbf{\textit{q}} ^{\mbox{\tiny L}} n_\textbf{1}^{\mbox{\tiny L}} ( n_\textbf{2}^{\mbox{\tiny L}} + n_\textbf{3}^{\mbox{\tiny L}} ) ]\,, 
 \end{split}\\
  \begin{split}
  	\label{StC}
 	\mbox{St}^{^{\rm LU}}(\textbf{\textit{q}} ,t)=&  \frac{\pi}{4} \int  d\textbf{\textit{q}}_1 d\textbf{\textit{q}}_2 d\textbf{\textit{q}}_3 \, \delta({\textbf{\textit{q}}+\textbf{\textit{q}}_1-\textbf{\textit{q}}_2-\textbf{\textit{q}}_3})\\  %\label{StD}
  &\times	 \delta (\Omega_\textbf{\textit{q}}^{^{\rm U}} +\Omega_1^{^{\rm U}}-\Omega_2^{^{\rm L}} -\Omega_3^{^{\rm L}}) \, 
 	|T^{^{\rm LU}}_{{\textbf{\textit{q}} 1\, \textbf{2} \textbf{3}}}|^2  \, \\%\label{StD}
  &\times	 [n_\textbf{2}^{\mbox{\tiny U}}  n_\textbf{3}^{\mbox{\tiny U}} (n_\textbf{\textit{q}}^{\mbox{\tiny L}}  + n_\textbf{1}^{\mbox{\tiny L}} )- n_\textbf{\textit{q}}^{\mbox{\tiny L}} n_{\bm 1}^{\mbox{\tiny L}} ( n_\textbf{2}^{\mbox{\tiny U}}  + n_\textbf{3}^{\mbox{\tiny U}} ) ]\, .
  \end{split}	
 \end{align}
\end{subequations} 

%=================================================================================================
\section{\label{s:model} Double accumulation of hybrid bosons} 
A full consistent theoretical description of the bottleneck accumulation process of hybrid magneto-elastic bosons, based on the kinetic equation\,(\ref{St}) for the actual YIG frequency spectra and the interaction amplitudes, is beyond the scope of this article. 
In \Sec{ss:dif}, for a quantitative understanding of this problem, we restrict ourselves to the simplifying assumption of axial symmetry of the problem in the vicinities of the frequency minima at $\B q=\pm \B q_0$. Then, in \Sec{s:2D}, we turn to a two-dimensional description of the problem, considering caustics in the quasiparticle propagation which results in sharp anisotropic beams.
%-------------------------------------------------------------------------------------------------
\subsection{\label{ss:dif} One-dimensional differential model}

Denoting $\B \kappa_\pm=\B q\mp \B q_0$ we introduce a one-dimensional approximation of the L- and U-MEM densities
 \begin{subequations}
  	\begin{equation}\label{not}
  	N^{^{ \rm L}}_{\kappa_\pm}= 4\pi \kappa_\pm^2 n^{\mbox{\tiny L}}_{\kappa_\pm}\,, \ 	N\Sp{U}_{\kappa_\pm}= 4\pi \kappa_\pm^2 n^{\mbox{\tiny U}}_{\kappa_\pm}\ .
  	\end{equation}
To simplify further the appearance of the equations, we will ignore the difference between the $\pm \B q_0$ frequency minima and omit the symbol $``\pm"$. This simplification does not affect the qualitative picture of the described phenomena.  
  	
Now we can present the kinetic equation\,(\ref{KE}) as follows 
  	 \begin{align} \label{KE-L}
 \frac{\partial N^{^{\rm L}}_\kappa}{ \partial t} & =  4\pi  \kappa^2 \, [\mbox{St}^{^{\rm LL}}(  \kappa   , t)+ \mbox{St}^{^{\rm LU}}(  \kappa, t)] \,, \\ \label{KE-U}
 \frac{\partial N^{^{\rm U}}_ \kappa}{ \partial t} & =  4\pi \kappa^2 \, [\mbox{St}^{^{\rm UU}}( \kappa, t)- \mbox{St}^{^{\rm LU}}( \kappa, t)] \ .
  \end{align}
\end{subequations}  

%-------------------------------------------------------------------------------------------------
\subsection{\label{ss:scattering} Estimates of the flux and the L$\to$U transformation rate}

The collision terms $\mbox{St}^{^{\rm LL}}(\kappa ,t)$ and $\mbox{St}^{^{\rm UU}}(\kappa ,t)$ preserve the total number of quasiparticle in the L-MEM and U-MEM, respectively. Therefore, they may be represented in a divergent form, for instance,
\begin{subequations}\label{10}\begin{equation}\label{10A}
\mbox{St}^{^{\rm LL}}( \kappa ,t)= d \mu_\kappa ^{\mbox{\tiny L}}/ d \kappa\,,
\end{equation}
    where $\mu_\kappa\Sp L$ is the L-MEM particle flux towards small wavenumbers. Together with Eq.\,(\ref{10A}), this gives
    \begin{equation} \label{10B}
    \mu_\kappa\Sp L= 4\pi\int^\kappa \kappa^2 \mbox{St}^{^{\rm LL}}(\kappa') d\kappa'\ .
    \end{equation} 
Now, using Eq.\,(\ref{StB}), we obtain an estimate for the flux $\mu_q ^{\mbox{\tiny L}}$:
 \begin{equation} \label{10C}
\mu_\kappa    \Sp L \simeq \kappa^3_\times  \big(T_\kappa \Sp{LL}\big)^2 \big(N_\kappa\Sp{L}\big)^3 \big / \delta\omega_\kappa\ .
\end{equation} 
Here, $\delta \omega_\kappa \equiv  \kappa^2  \big[  d^2 \omega_q/ 2 (d q)^2\big]_{q=q_0} $ and $T_\kappa\Sp{LL}= T_0 \cos^4 \varphi_\kappa$, with $T_0=T_{\B q_0\B q_0, \B q_0\B q_0}$ and $\cos \varphi_\kappa$ given by \Eq{3b} were $\B q=\B q_0+\B \kappa$. 
A similar estimate of the collision integral $\mbox{St}^{^{\rm LU}}_\kappa = \mbox{St}^{^{\rm LU}}(\kappa) $, \Eq{StC}, takes the form
\begin{equation}\label{10D}
 \mbox{St}^{^{\rm LU}}_\kappa  \simeq \kappa ^2_\times \big(T^{^{\rm LU}}_\kappa\big)^2\, N_\kappa^{^{\rm L}} N_\kappa ^{^{\rm U}}\big ( N_\kappa^{^{\rm U}} - N_\kappa^{^{\rm L}}\big ) \big / \delta\omega_\kappa 
\end{equation}\end{subequations}
with $T_\kappa\Sp{LU}= T_0 \cos^2 \varphi_\kappa\sin^2 \varphi_\kappa$. It can be shown that conservation laws of frequency and momentum fully forbid this scattering process when the frequency of the lower mode becomes smaller than the lowest possible frequency of the upper mode, i.e., the BEC frequency $\omega_0$. It means that the estimate\,\eqref{10D} is valid if $\Omega^{^{\rm L}}_{\kappa} > \min \Omega^{^{\rm U}}_{\kappa}=\omega_0$. Otherwise, $\mbox{St}^{^{\rm LU}}_\kappa =0$. Note that \Eqs{10C} and \eqref{10D} are more accurate than the preliminary estimates given in \Refn{arc-1}.

\begin{figure*}
	\includegraphics[width=1\textwidth]{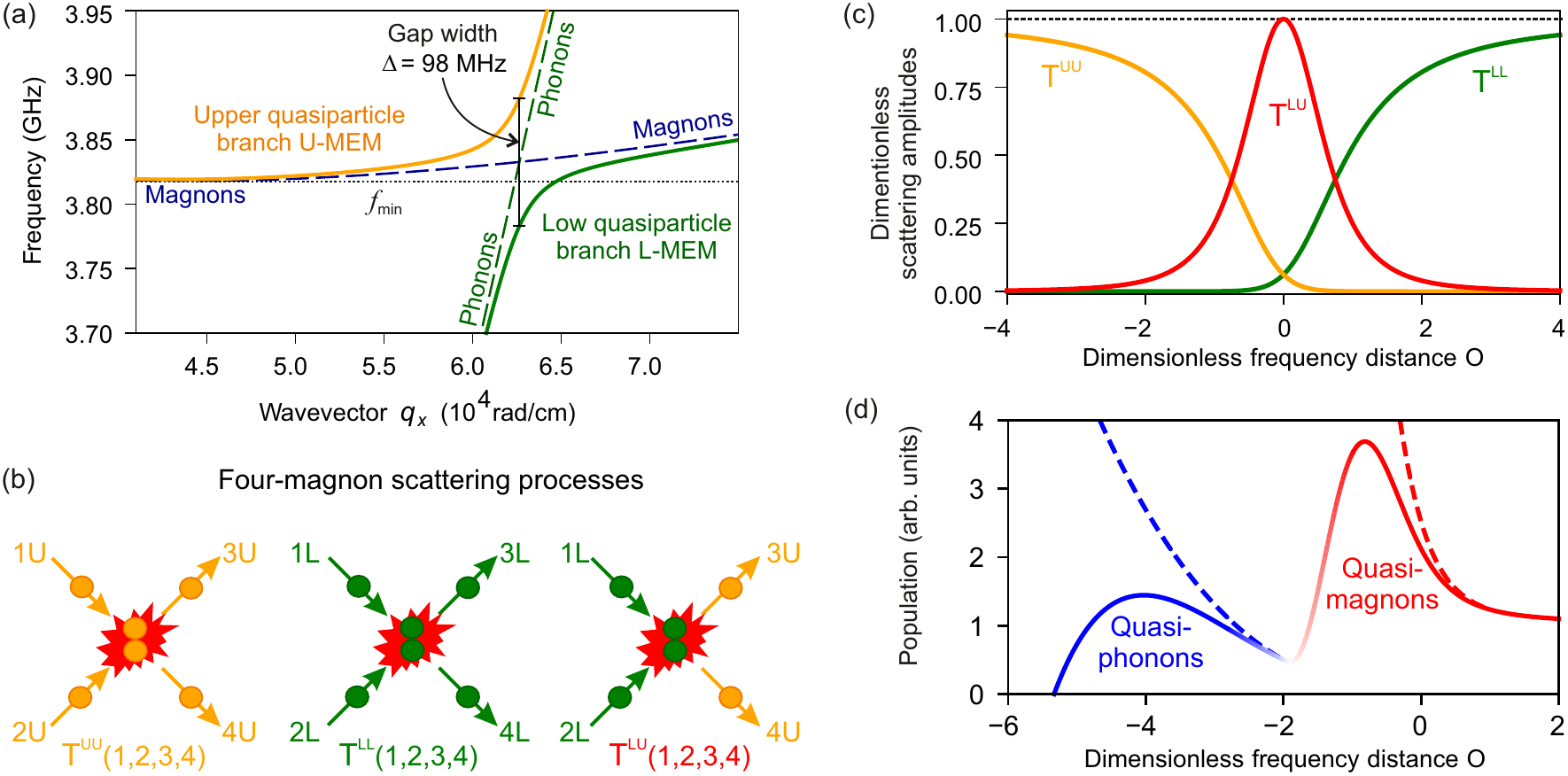}
	\caption{
		\label{model} 
		Panel (a) shows the calculated magnon-phonon spectrum near the hybridization region. Panel (b) gives the schematic representations of the scattering and cross-scattering four-particle processes in the hybridization region. Panel (c) presents squares of interaction amplitudes $T^{^{\rm UU}}_q$, $\ T^{^{\rm LL}}_q$  and $4 T^{^{\rm LU}}_q$, normalized by $T_q$, vs the dimensionless distance from the hybridization crossover $O_q$.  Panel (d): Solutions of balance equations in the cross-section area $-2<O_\kappa<2$, (solid red line) and below the BEC frequency $-2>O_\kappa$, solid blue line. Red and blue dashed lines: Particle profiles $N\Sp L_\kappa$ for constant particle fluxes $\mu\Sp L_\kappa=$const.
	}
\end{figure*}

Now, in the stationary case the kinetic equation\,(\ref{KE-L}) for the lower mode can be written as follows:
\begin{equation}\label{flux}
\frac{d \mu\Sp L_\kappa}{d \kappa}+ \mbox{St}\Sp{  LU}_\kappa =0\ .
\end{equation}  

%-------------------------------------------------------------------------------------------------
\subsection{\label{ss:sol-1} Rate equation above the BEC frequency}
  
Using estimates\,\eqref{10C} and \eqref{10D} we can rewrite the rate \Eq{flux} as
\begin{align}
  	\begin{split}\label{bal-1}
& \frac{d}{d \kappa}\big [\kappa_\times^3\cos^8 \varphi_\kappa \big(\C N_\kappa\Sp L\big)^3\big]\\ =& 3\, a \big(\C N_\kappa\Sp L\big)^2 \kappa^2 \cos^4 \varphi_\kappa\sin^4 \varphi_\kappa\ .
	\end{split}  
\end{align}
Here, $\C N_\kappa\Sp L=  N_\kappa\Sp L/N_+\Sp L$ is the dimensionless L-MEM density, normalized by the density $N_+\Sp L$ at a sufficiently large positive $O_\kappa $, say at $O_\kappa=3$. In \Eq{10D}, we assume for simplicity's sake that $N_\kappa\Sp L\gg N_\kappa\Sp U$ and prescribe $N_\kappa\Sp U$ as $\kappa^2 N_+\Sp L/\kappa_\times^2$ according to \Eq{not} with $ n_\kappa^{\mbox{\tiny U}}=$const. The dimensionless parameter $a\sim 1$ combines all uncontrolled parameters in our estimates.
 
The ordinary differential equation\,\ref{{bal-1}} can be solved with the boundary condition $\C N_\kappa\Sp L=1$ for $\kappa\to \infty$ giving the relative MEMs population in the hybridization region (above the BEC frequency and thus with a plus-symbol)
  \begin{align}
  \C N  _\kappa \Sp {+,L}=\frac 1{\cos^{8/3}\varphi_\kappa }\Big [1- a \int \limits ^\infty _{\kappa/\kappa_\times }\frac {x^2 \sin^4 \varphi_x d x}{cos^{4/3}\varphi_x}\Big] \ .
  \end{align}
In Fig.\,\ref{model}(d)  we plot $\C N\Sp {+,L}_\kappa$ in red as a function of the dimensionless distance to the crossover $O_\kappa$ defined by \Eq{3b}. For concreteness we took the position of the frequency minimum $\kappa=0$ as $O_0=-2$. We see a sharp peak of $\C N\Sp {+,L}_O$ demonstrating the bottleneck accumulation of quasi-magnons in the hybridization region above the BEC frequency around $O\approx -1$. This peak is a result of the completion of two processes. The first one is the quasiparticle flux towards lower frequencies (negative $O$), which on its own leads to an infinite growth of $N\Sp {+,L}_O$, shown by the red dashed line in Fig.\,\ref{model}(d). This growth is caused by a decrease in the intrinsic LL nonlinearity [see plot of $T_\kappa\Sp{LL}$ in Fig.\,\ref{model}(d)], which has to be compensated by an increase in $N\Sp {+,L}_O$ to ensure a constant particle flux. This increase of $N\Sp {+,L}_O$ is limited by the second process: the intermodal L$\to$U particle flux, provided by the St$\Sp{LU}_\kappa$ collision integral\,\eqref{10D}. For frequencies of the lower mode  $\Omega^{^{\rm L}}_\kappa <\omega_0$  (i.e., below the BEC frequency) St$\Sp{LU}_\kappa$ becomes zero (by the conservation laws of frequency and momentum) and growth of $N\Sp L_O$ reappears, see blue dashed line in Fig.\,\ref{model}(d). We consider this effect in \Sec{ss:sol-2}. 

%-------------------------------------------------------------------------------------------------  
\subsection{\label{ss:sol-2} Rate equation below the BEC frequency}

Below the BEC frequency we have St$\Sp{LU}_\kappa=0$ (see \eq{flux}) and we have to account for another dissipation mechanism able to suppress the infinite growth of  $\C N\Sp L_\kappa$. The simplest option is a small linear damping term $\gamma \C N\Sp L_\kappa$ originating from three-magnon scattering processes, magnon-phonon interaction or other processes. With this term instead of \Eq{flux} we have:
 \begin{equation}\label{flux1}
 \frac{d \mu^{\mbox{\tiny  L}}_\kappa}{d \kappa}+ \gamma \C N\Sp{LU}_\kappa =0\ .
  \end{equation} 
 Now, using an estimate\,\eqref{10C} we can rewrite rate \Eq{flux1}  similarly to \Eq{bal-1}
 \begin{align}
 	\begin{split}\label{bal-2}
 & \frac{d}{d \kappa}\big [\kappa_\times \cos^8 \varphi_\kappa \big(\C N_\kappa\Sp L\big)^3\big]  +  \frac 32  b \, \C N_\kappa\Sp L=0  \ .
    \end{split}  
 \end{align}
The dimensionless parameter $b\ll 1$ combines all the uncontrolled parameters in our estimates including a small damping term. 
 
The solution of the ordinary differential equation\,(\ref{bal-2}) with proper boundary conditions at the BEC frequency (i.e. for $\kappa=0$) provides the relative population of the L-MEM branch for smaller frequencies (minus-symbol) 
 \begin{align}
 \C N _\kappa \Sp{$-$,L}= \C N _0 \Sp{+,L}\frac {\cos^{8/3}\varphi_0 }{\cos^{8/3}\varphi_\kappa } \Big [1- b \int \limits ^0 _{\kappa/\kappa_\times }\frac { d x}{cos^{4/3}\varphi_x}\Big]^{1/2}\ .
 \end{align}
In Fig.\,\ref{model}(d) we plot $\C N\Sp{$-$,L}$ by a solid blue line as a function of the dimensionless distance to the crossover $O_\kappa$ below the BEC frequency (in our case for $O<-2$). For concreteness we took $b=0.01$. We see a second sharp peak of $\C N\Sp L_O$ demonstrating the bottleneck accumulation of quasi-phonons much below the BEC frequency around $O\approx -4$. 

%=================================================================================================
\vspace{-3mm}
\section{\label{s:2D}Two-dimensional model: \\ caustics and anisotropic beams}
\vspace{-3mm}

	\begin{figure*}
		\centering
		\includegraphics{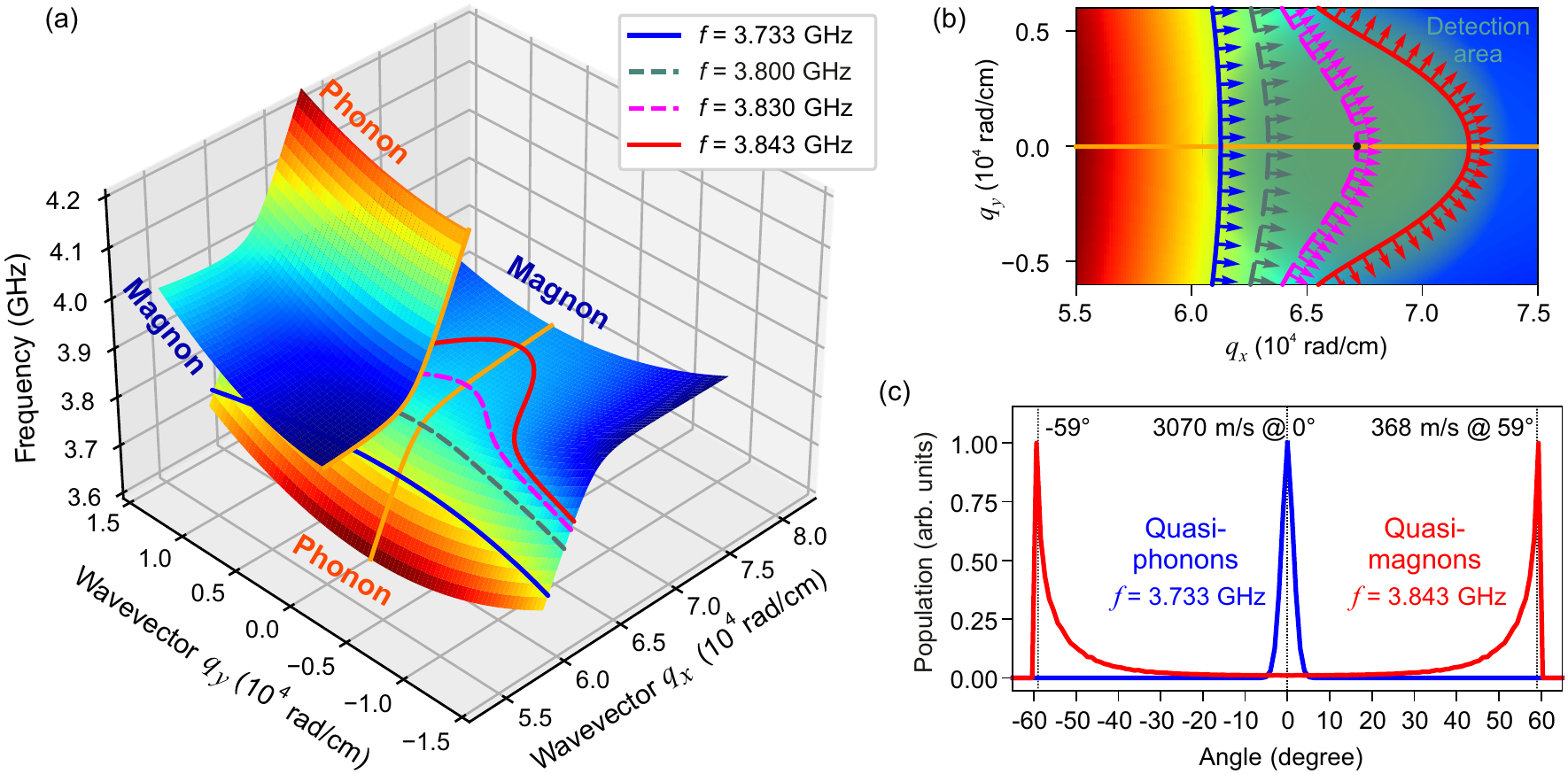}
		\caption{ 
			\label{caustics} 
			 The graph in panel (a) shows the two-dimensional dispersion of magneto-elastic quasiparticles at the magnon-phonon crossing. The color scale symbolizes the kind of the particle. dark red stands for a pure phonon state, while blue stands for a pure magnon state. Two continuous orange lines mark the dispersion branches with the perpendicular wavevector $k_\mathrm{y} = 0$ rad/cm. The red, purple and blue lines are selected isofrequency curves. (b) This panel shows an intensity map of the lower part of the magnon-phonon dispersion shown in panel (a). The arrows indicate the directions of quasiparticle propagation in different sections of the isofrequency curves, and , by definition of the group velocity, they are always perpendicular to the isofrequency curves. The green circle symbolizes the $3\sigma$ resolution of the measurement setup in wavevector space. Panel (c) shows the quasiparticle population along the isofrequency curve in dependence of the propagation angle $\alpha$. 
			 It becomes obvious that each of the shown isofrequency curves has preferred angles and velocities of the quasiparticles populating it. These angles increase with growing frequency. The propagation angle of the blue curve is nearly 0$^{\circ}$, while the red curve has two angles of $\pm 59^{\circ}$.}
	 
	\end{figure*} 

Above, we formulated a one-dimensional model of the bottleneck accumulation of hybrid magneto-elastic quasiparticles during the thermalization of parametrically pumped  magnons. 
The model predicts the existence of two regions of such accumulation (slow quasi-magnons and fast quasi-phonons) with frequencies above and below the BEC frequency, $f\sb{slow}> f_0 > f\sb{fast}$,  exactly as observed in the experiment.

However, we are unable to determine the spectral positions of these peaks quantitatively. The problem is that instead of integration over the wavevector space we used local estimations \eq{10C} and \eq{10D} of the collision integral~(\ref{StA}). This approximation works more or less reasonably well, in hydrodynamic turbulence theory with a large scale-invariant interval in wavevector space and smooth dependence of interaction amplitudes, as in Ref.\,\cite{Lvov-Pomyalov2018}.
However, this is not the case in our problem: the relevant interaction amplitudes demonstrate a sharp $q$-dependence in the vicinity of the crossover region. Thus, a quantitative description of the problem require an explicit solution of the rate equation with the actual form (\ref{StA}) of the collision integral and realistic boundary conditions.

Moreover, the suggested one-dimensional model leaves the question unanswered about the angular dependence of the propagation directions of the slow and fast quasiparticles, which should be the subject of a full three-dimensional theory. Such a theory must take into account the strongly anisotropic frequency spectrum of magnons and requires knowledge of the three-dimensional population of the L-MEM branch above the crossing frequency $f_\times$, and the U-MEM branch from $f_0$ to $f_\times$ and higher. Neither do we know these distributions experimentally nor theoretically. In this situation, we can only make the simplest assumptions about the mentioned distribution functions and compare the obtained results with the experiment. 
 
Figure\,\ref{caustics}(a) shows two-dimensional frequency spectra of U- and L-MEMs in the vicinity of the magnon-phonon hybridization area. With the solid blue line we plotted here the isofrequency curve of fast quasi-phonons at $3720\,$MHz.  Its shape looks like a segment of a perfect circle formed by the intersection of the isofrequency plane with the phonon dispersion cone, only slightly distorted by hybridization with the magnon spectrum.  The red solid line shows the isofrequency curve of slow quasi-magnons at $3843\,$MHz.  This curve deviates significantly from the circular shape due to the influence of strongly anisotropic magnon dispersion. To more clearly demonstrate the transition between these two curves, we also plotted two intermediate isofrequency curves at $3800\,$MHz and $3830\,$MHz as dashed lines. The same four curves are also represented in the $(q_x,q_y)$-plane in Fig.\,\ref{caustics}(b). They are complemented by arrows indicating the directions of group velocities. It is clear that the L-MEM quasiparticles belonging to these isofrequency curves have a different primary direction of propagation.  

Obviously, the sensitivity of the BLS detection of quasiparticles $S_{\B q}$ is limited in the $(q_x,q_y)$-space. In our experiment, the center of the area of maximum sensitivity corresponds to the position of the small black circle in Fig.\,\ref{caustics}(b). When moving away from it, the sensitivity decreases approximately tantamount to a Gaussian curve with a standard deviation of $\sigma \sim 10^3\,$rad/cm. A qualitative estimate of the observational region in which we can register quasiparticles is shown as a circle-shaped green area. 
We see that it includes some part of the blue line shown in Fig.\,\ref{caustics}(b). The group velocities of all corresponding quasi-phonons are almost directed along $q_x$. On the other hand, the quasi-magnons belonging to the red isofrequency curve in the green area of observation can be divided into two groups with positive and negative $q_y$, above and below the horizontal orange solid line. A substantial fraction of the upper part of the isofrequency curve around the inflection point has almost equally directed group velocities pointing upward at an angle of about $60^\circ$, causing the formation of a caustic in the propagating quasiparticle beam. Similarly, the lower part of the isofrequency curve with $q_y<0$ is responsible for the formation of the second caustic beam propagating at an angle $\alpha=-60^\circ$ with respect to the horizontal line $q_y=0$.  

To qualitatively characterize the distribution of group velocities $\B v\sb{gr}\sp{\tiny L} (\B q)$ within each L-MEM quasiparticle group registered in our experiments, we must find the product $ S_{\B q} n_{\B q}\sp{\tiny L} \B v\sb{gr}\sp{\tiny L} S (\B q)$ along the corresponding part of the dispersion curve. Unfortunately, we do not know the L-MEM population $n_{\B q}\sp{\tiny L}$ and can only assume for simplicity's sake that $n_{\B q}\sp{\tiny L}=\mathrm{const}$ in the actual area. The resulting distributions of quasiparticles vs. the group velocity angle are shown in Fig.\,\ref{caustics}(c), where all plots are normalized to their maximum value for better comparison. We see three sharp peaks: a peak of fast quasi-phonons with $\alpha=0^\circ$ and two peaks of slow quasi-magnons with $\alpha\approx \pm 59^\circ$. This perfectly resembles the experimentally measured direction of propagation of slow and fast quasiparticles. By extracting the velocities for the points on the isofrequency curves matching to propagation angles $\alpha=0^\circ$ and $\pm 59^{\circ}$, it becomes possible to calculate the average propagation velocity values $v_\mathrm{gr}^\mathrm{theor}$ for both propagation angles. The results are $v_\mathrm{gr}^\mathrm{theor}(0^{\circ}, 3720\,\mathrm{MHz})$=3215\,m/s and $v_\mathrm{gr}^\mathrm{theor}(59^{\circ}, 3843\,\mathrm{MHz})$=360\,m/s, which are in good agreement with our experimental data (see Table\,\ref{t:1}). 

 \begin{table}[b]
 	\begin{tabular}{c||c|c|c|c}
 		                                                                                 &     Fast      &     Slow      &  BEC   & Crossing \\
 		                                                                                 & quasi-phonons & quasi-magnons & minima &  point   \\ \hline\hline
 		                         $ f $ (MHz)                                             &     3720      &     3843      &  3819  &   3833   \\
 		$q_x  \big( \frac{\mbox{rad}}{\mbox{cm}} \big)\textcolor{white}{\Big |}$                    &    61238      &    67670      & 43360  &  62620   \\
 		$q_y  \big( \frac{\mbox{rad}}{\mbox{cm}} \big)\textcolor{white}{\Big |}$                    &        0      &  $\pm 4600$   &     0  &      0   \\
   $v_\mathrm{gr}^\mathrm{theor}  \big( \frac{\mbox{m}}{\mbox{s}} \big)\textcolor{white}{\Big |}$   &     3215      &     360       &     0  &      -   \\
   $v_\mathrm{gr}^\mathrm{exp}  \big( \frac{\mbox{m}}{\mbox{s}} \big)\textcolor{white}{\Big |}$     &     3070      &     368       &     0  &      -   \\
 	\end{tabular}
 	\caption{\label{t:1} 
 	Characteristic values of the problem determined from the two-dimensional model. 
 	}
 \end{table}

One should distinguish a somewhat different nature of the directed quasiparticle fluxes for angles of $59^\circ$ and $0^\circ$. In the first case, the isofrequency curve has two inflections at the points ($q_x \approx 67670 \, \mathrm{rad/cm}, q_y \approx \pm 4600 \, \mathrm{rad/cm}$); in their vicinity, canonical caustic patterns are formed \citep{Veerakumar2006}. Such caustics are protected from diffraction broadening and have a stable transverse aperture, which can be of subwavelength size \citep{Serga2010}. In the second case, the isofrequency curve has no inflection points and, thus, a conventional weakly divergent beam is formed by quasiparticles with approximately co-directed group velocities \cite{Heussner2020}. 

It should be noted that any isofrequency curve lying between the blue and red curves, being populated, should form  either focused or caustic beams at some propagation angles in the range from 0 to $\pm 59^{\circ}$. In our experiment, however, we do not observe quasiparticle propagation between these two boundary cases. This fact convincingly confirms the model of double accumulation of magneto-elastic bosons in the magnon-phonon hybridization region.

%=================================================================================================
\vspace{-3mm}
\section{\label{s:discussion} Discussion and summary} 
\vspace{-3mm}

We experimentally showed the appearance of two groups of hybrid magneto-elastic modes at the bottom of the spectrum of a parametrically overpopulated magnon gas in an in-plane magnetized magnetic film. These two groups form spatially separated beams with different group velocities. The first---``slow''---group propagates with velocity $ v\sb{slow}\approx 368\,$m/s under the angles $\alpha\sb{slow} \approx \pm 59^\circ$  with respect to the external magnetic field $\B H\| \hat {\B x}$.
The second---``fast''---group appears later  and propagates with velocity $ v\sb{fast}\approx 3070\,$m/s along $\hat {\B x}$: $ \B v\sb{fast}\| \B H\| \hat {\B x}$.
  
We formulated a simple one-dimensional model of the bottleneck accumulation of hybrid MEMs during the process of parametrically pumped magnons evolution toward their Bose-Einstein condensation. The model predicts two accumulation areas (slow quasi-magnons and fast quasi-phonons) with the frequencies above and below the BEC frequency, $f\sb{slow}> f_0> f\sb{fast}$, exactly as observed in experiments. We consider the qualitative agreement of the experimental results with the simple analytical model as a strong evidence that our one-dimensional model grasps adequately the basic physics of the bottleneck accumulation phenomenon in the frequency domain.
  
To explain the reason for the observed slow and fast quasiparticle propagation in narrow angular intervals around $\alpha\sb{slow}\sp{exp}\approx \pm 59^\circ$ and $\alpha\sb{fast}\sp{exp}\approx 0^\circ$ we considered in \Sec{s:2D} two-dimensional frequency spectra and found well-defined regions with almost the same group velocities--caustics--around the inflection points in the BLS registered part of the $\B q$-space. Using only this knowledge, we theoretically found propagation angles $\alpha\sb{slow}\sp{theor}= \pm 59^\circ$ and $\alpha\sb{fast}\sp{theor}= 0^\circ$  [see \Fig{caustics}(c)], which perfectly agree  with  corresponding experimental values $\alpha\sb{slow}\sp{exp}\approx \pm 59^\circ$ and $\alpha\sb{fast}\sp{exp}\approx 0^\circ$. 

We consider the discovered double accumulation of magnetoelastic modes with nonzero group velocities as a promising effect for applications in future magnon spintronic devices. To determine the specific field of such applications, one should clarify the degree of real spectral localization of each of these quasiparticle groups and, consequently, the degree of their coherence. We believe that the answer to this question can be obtained by interference experiments with accumulated quasiparticles. 

%================================================================================================
\vspace{-3mm}
\section*{Acknowledgments}
\vspace{-3mm}

 Financial support by the European Research Council within the Advanced Grant 694709 SuperMagnonics -- ``Supercurrents of Magnon Condensates for Advanced Magnonics'' as well as financial support by the Deutsche Forschungsgemeinschaft (DFG, German Research Foundation) through the Collaborative Research Center ``Spin+X: Spin in its collective environment'' TRR -- 173 -- 268565370 (project B04) is gratefully acknowledged. We thank G. A. Melkov for fruitful discussion.

\end{document}